\pdfoutput=1

\documentclass[11pt,twoside,a4paper,cmspaper,final,collab]{cms-tdr}
\def\svnVersion{167921}\def\cmsCernNo{2012-174}\def\cmsCernDate{\today}\def\cmsMessage{Submitted to the Journal of High Energy Physics}

\begin{document}\cmsNoteHeader{HIG-12-009}

\hyphenation{had-ron-i-za-tion}
\hyphenation{cal-or-i-me-ter}
\hyphenation{de-vices}

\RCS$Revision: 134832 $
\RCS$HeadURL: svn+ssh://svn.cern.ch/reps/tdr2/papers/HIG-12-009/trunk/HIG-12-009.tex $
\RCS$Id: HIG-12-009.tex 134832 2012-07-04 21:43:25Z alverson $
\providecommand{\met}{\MET}
\providecommand{\Wjets}{\ensuremath{\PW+\text{jets}}}
\providecommand{\Zjets}{\ensuremath{\cPZ+\text{jets}}}

\newcommand{\FPObsA}{{110}}
\newcommand{\FPObsB}{{194}}
\newcommand{\FPObsAA}{{110}}
\newcommand{\FPObsBB}{{188}}
\newcommand{\FPObsGapOneAA}{{124.5}}
\newcommand{\FPObsGapOneBB}{{127}}
\newcommand{\FPObsGapTwoAA}{{147.5}}
\newcommand{\FPObsGapTwoBB}{{155}}
\newcommand{\FPMinLocalP}{{0.003}}
\newcommand{\FPMaxLocalZ}{{2.5}}
\newcommand{\FPMaxZmass}{{125}}

\cmsNoteHeader{HIG-12-009} % This is over-written in the CMS environment: useful as preprint no. for export versions
\title{Search for a fermiophobic Higgs boson in pp collisions at $\sqrt{s}=7$\TeV}

\date{\today}

\abstract{
Combined results are reported from searches for a fermiophobic Higgs
boson in the $\gamma\gamma$, WW, and ZZ decay modes in proton-proton collisions at
 $\sqrt{s}=7$\TeV. The explored Higgs boson mass range is 110--300\GeV.
 The data sample corresponds to an integrated luminosity of 4.9--5.1\fbinv.
 A fermiophobic Higgs boson is excluded at 95\% confidence level in the mass
 range 110--194\GeV, and at 99\% confidence level in the mass ranges 110--124.5\GeV,
 127--147.5\GeV, and 155--180\GeV.
}

\hypersetup{%
pdfauthor={CMS Collaboration},%
pdftitle={Search for a fermiophobic Higgs boson in pp collisions at sqrt(s)=7 TeV},%
pdfsubject={CMS},%
pdfkeywords={CMS, physics, Higgs}}

\maketitle %maketitle comes after all the front information has been supplied

\newcommand{\CLS}{\ensuremath{CL_\mathrm{S}}\xspace}
\newcommand{\gamgam}{\ensuremath{{\gamma\gamma}}\xspace}
\newcommand{\ptgg}{\ensuremath{p_{\mathrm{T}}^{\gamma\gamma}}\xspace}
\newcommand{\mgg}{\ensuremath{m_{\gamma\gamma}}\xspace}
\newcommand{\piTgg}{\ensuremath{{\pi_\mathrm{T}^{\gamma\gamma}}}\xspace}
\newcommand{\ygg}{\ensuremath{{y_{\gamma\gamma}}}\xspace}
\newcommand{\cosThetaGG}{\ensuremath{{\cos\theta_{\gamma\gamma}}}\xspace}
\newcommand{\piT}{\ensuremath{{\pi_\mathrm{T}}}\xspace}
\newcommand{\HqT}{{\textsc{HqT}}\xspace}
\newcommand{\mH}{\ensuremath{m_{\mathrm{H}}\xspace}}
\section{Introduction}
\label{sec:intro}

In the standard model (SM), the electroweak symmetry-breaking takes place through the Higgs mechanism
in which a complex scalar doublet with a non-zero vacuum expectation value is introduced and
the existence of one scalar particle, the Higgs boson, is predicted~\cite{Higgs1,Higgs2}.
The gauge bosons derive their masses from the additional degrees of freedom gained from the symmetry breaking,
and the fermions acquire mass through the direct interaction with the Higgs field itself.
It is possible that the mechanism that generates the fermion masses is independent of the Higgs boson.
Such a Higgs boson is usually referred to as fermiophobic (FP)~\cite{akeroyd95, akeroyd96}.
Its decay to W and Z bosons proceeds as in the SM, while the decay to photons proceeds
via W loops, i.e. decays to photons via fermion loops are excluded by the model.
Since decays to \bbbar and
$\tau\tau$ are forbidden at tree-level, the branching fraction for a low mass FP Higgs boson ($\mH
\approx 120\GeV$) to decay to two vector bosons or two photons is enhanced by an order of magnitude with respect to
the SM~\cite{gunion89, akeroyd98, mele_2010}.
Previous searches at LEP~\cite{LepFP}, the Tevatron~\cite{TevatronFP}, and the LHC~\cite{AtlasFP}
rule out an FP Higgs boson lighter than 121\GeV at 95\% confidence level (CL).

In this letter we report on a search for an FP Higgs boson in the mass
range 110--300\GeV, in the $\gamgam$, WW, and ZZ decay modes
in proton-proton collisions at a centre-of-mass energy of 7\TeV
with data collected in 2011 by the Compact Muon Solenoid (CMS) detector at the LHC.
The production of the FP Higgs boson is by vector boson fusion (VBF) and associated production with a vector boson (VH).
With respect to the SM, the signal is suppressed by an order of magnitude for $\mH > 150\GeV$, is comparable for $\mH \approx 130\GeV$, and is enhanced by an order of magnitude for $\mH \approx 110\GeV$.
While for the cases of WW and ZZ the search relies on a re-interpretation of the standard model Higgs boson searches, for the $\gamgam$ final states a dedicated analysis has been put in place.
The descriptions of the analyses emphasise the sub-channels and
techniques not previously described in recent publications of SM
analyses~\cite{Hgamgam, HWW, HZZ4l, HZZ2l2nu, HZZ2l2q, HZZ2l2tau}, namely:
the lepton tag and the use of a two-dimensional fit in the \gamgam decay channel, and the lepton tag in the WW decay channel.

\section{The CMS detector}
\label{sec:detector}

While the CMS detector is described in detail elsewhere~\cite{CMSdetector}, the
key components for this analysis are summarised here.

The central feature of the CMS apparatus is a superconducting solenoid, of 6~m internal diameter, providing a field of 3.8\unit{T}. Within the field volume are a silicon pixel and strip tracker, a crystal electromagnetic calorimeter (ECAL) and a brass/scintillator hadron calorimeter (HCAL).
Muons are measured in gas-ionisation detectors embedded in the steel return yoke. Extensive forward calorimetry complements the coverage provided by the barrel and endcap detectors.

The inner tracker measures charged particles within the pseudorapidity range $\abs{\eta}< 2.5$. It consists of 1440 silicon pixel and 15\,148 silicon strip detector modules. %It provides a \pt resolution of about 1.5\% for 100\GeV particles.

The ECAL consists of two parts: the barrel which covers the
pseudorapidity $\eta$ range $|\eta|<1.48$, and the endcaps
covering the range $1.48<|\eta|<3.0$,  where $\eta=-\ln{[\tan{(\theta/2)}]}$,
and $\theta$ is the polar angle of the anticlockwise trajectory of a particle
with respect to the beam direction.
When used for detector positions the trajectory is assumed to originate at the
nominal interaction point, corresponding to the coordinate system origin.
The ECAL consists of lead tungstate crystals
arranged in a quasi-projective geometry.
In the barrel region the front
face of the crystal is approximately $ 22 \times 22\unit{mm}^2$, corresponding to
a granularity of $\Delta\eta\times\Delta\phi=0.0174\times 0.0174$.
In the endcap the front face of the crystals is approximately $ 29
\times 29\unit{mm}^2$.

In the region $\vert \eta \vert< 1.74$, the HCAL cells have a
granularity of $\Delta\eta\times\Delta\phi=0.087\times 0.087$,
and for $\vert \eta \vert< 1.48$,  map onto arrays of $5 \times 5$
crystals in ECAL
to form calorimeter towers projecting radially outwards from
the nominal interaction point.
At larger values of $\vert \eta \vert$, the size of the towers
increases and the matching ECAL arrays contain fewer crystals.
Within each tower, the energy deposits in ECAL and HCAL cells are
summed to define the calorimeter tower energies, subsequently used to provide the energies and directions of hadronic jets.
A quartz-fibre Cherenkov calorimeter extends the coverage to $|\eta| < 5.0$.

\section{Search channels}
\label{sec:classes}

The search is performed using three decay modes of the FP Higgs boson.
In the analysis of the \gamgam and the WW decay modes, characteristic
signatures of FP Higgs boson production via VBF and VH
are exploited to select events and suppress background:
the two forward jets produced by the scattered quarks in VBF production (dijet tag)
and isolated charged leptons (electrons or muons) from decays of the vector bosons in VH production (lepton tag).

In the \gamgam decay mode, events with two isolated and high transverse momentum (\pt)
photons are selected.
Seven event classes are defined:
three of diphoton events which additionally require at least a pair of
jets, an isolated muon, or an isolated electron, respectively,
and four comprising the remaining diphoton events which are subdivided
according to the photon shower shape and position in the detector~\cite{Hgamgam}.

In the WW decay channels, the events are characterised by the presence of two opposite sign,
isolated high \pt leptons from the W decay,
together with large missing transverse energy ($\ETmiss$) due to undetected neutrinos.
One sub-channel is defined by additionally requiring the presence of two jets with the VBF topology~\cite{HWW},
and another by requiring the presence of a third isolated, high \pt lepton.
The selection requiring a third lepton is intended to select
Higgs boson production in association with a W boson (WH).
Final states with one or zero jets, each separated into events where
the leptons have the same flavour and those where there is an electron and
a muon, add a further four WW sub-channels, giving a total of six.

In the case of the ZZ final state, production mode signatures are not
exploited, and the results of SM Higgs searches ~\cite{HZZ4l,
HZZ2l2nu, HZZ2l2q, HZZ2l2tau}, comprising 19 sub-channels,
are simply re-interpreted from the different signal rates
expected in an FP Higgs boson model.

The final result is obtained from the combination of 32 mutually
exclusive sub-channels from the three decay modes, \gamgam, WW, ZZ, as summarised in Table~\ref{tab:chans}.
The luminosity calculation for the datasets used has been updated with
respect to that used in the SM Higgs production analysis of the same channels,
published in the references given in the table.

\begin{table}[b]
  \begin{center}
    \topcaption{Summary of analysis channels and sub-channels included in the combination.}
    \label{tab:chans}
    \begin{tabular}{l c c c c} \hline \hline
\multirow{2}{*}{Channel} & $\mH$ range & Sub- & Luminosity & \multirow{2}{*}{Reference} \\
                                       & (\GeVns) & channels & (\!\fbinv) & \\ \hline
      H$\to\gamgam$  & 110--150 & 4 & 5.1 & \cite{Hgamgam} \\
      H$\to\gamgam$ + dijet & 110--150 & 1 & 5.1 & \cite{Hgamgam} \\
      H$\to\gamgam$ + lepton & 110--150 & 2 & 5.1 & \\ \hline
      H$\to$WW$\to 2\ell 2\nu$ & 110--300 & 4 & 4.9 & \cite{HWW} \\
      H$\to$WW$\to 2\ell 2\nu$ + dijet & 110--300 & 1 & 4.9 &  \cite{HWW} \\
      H$\to$WW$\to 2\ell 2\nu$ + lepton & 110--300 & 1 & 4.9 & \\ \hline
      H$\to$ZZ$\to 4\ell$ & 110--300 & 3 & 5.0 & \cite{HZZ4l} \\
      H$\to$ZZ$\to 2\ell 2\nu$ & 250--300 & 2 & 5.0 & \cite{HZZ2l2nu}\\
      H$\to$ZZ$\to 2\ell 2$q & 130--165, 200--300 & 6 & 5.0 & \cite{HZZ2l2q}\\
      H$\to$ZZ$\to 2\ell 2\tau$ & 180--300 & 8 & 5.0 & \cite{HZZ2l2tau}\\
      \hline
    \end{tabular}
  \end{center}
\end{table}

The cross sections for the Higgs boson production mechanisms and decay branching fractions,
together with their uncertainties, are taken from Ref.~\cite{Dittmaier:2011ti} and are
derived from Refs.~\cite{Arnold:2008rz,Bolzoni:2010xr,Brein:2003wg,Ciccolini:2003jy,Djouadi:1997yw,hdecay2}.
The VBF production of the FP Higgs boson signal is simulated using the next-to-leading order matrix element generator \POWHEG 1.0 interfaced with \PYTHIA 6.4.24~\cite{Sjostrand:2006za} for parton showering and fragmentation.
The VH signal production channel is simulated with \PYTHIA.
Samples of Monte Carlo (MC) simulated events used in the analysis are passed through the \GEANTfour~\cite{Agostinelli:2002hh} model of the CMS detector and reconstructed with the same software as used for collision data.

\subsection{Diphoton \texorpdfstring{\gamgam}{gamma-gamma} decay mode}

In the \gamgam channel~\cite{Hgamgam},
two isolated photon candidates are required to be
within the ECAL fiducial region $|\eta| < 2.5$, excluding the
barrel-endcap transition region $1.44 < |\eta| < 1.57$.
The shape of electromagnetic shower is used to identify photons,
while track veto is used to exclude electrons.
Isolation is used to reject the background due to
electromagnetic showers originating in jets -- mainly due to single and multiple $\pi^{0}$s~\cite{Hgamgam}.
The isolation requirements are applied as a constant fraction of the candidate photon
\pt, effectively cutting harder on low \pt photons.
The $\RNINE$ variable, defined as the energy sum of 3x3 crystals centred on
the crystal with maximum energy deposit divided by the
total clustered energy, is used to distinguish photons of well measured energy.

\subsubsection{Dijet tag event class}
Candidate diphoton events for the dijet-tagged channel have the same
selection requirements as in the SM search~\cite{Hgamgam}.
In the events from the VBF production, the \pt of the Higgs boson is boosted giving enhanced asymmetries in the photon pair energies and hence favoring a lower threshold on one of the two photons.
The threshold requirements for this class are
$\pt^{\gamma}(1)~> 55\times\mgg/120$, and $\pt^{\gamma}(2)~> 25$\GeV.

For each event, hadronic jets are clustered from the reconstructed particles with the infrared and collinear safe anti-$k_\mathrm{t}$ algorithm~\cite{Cacciari:2008gp}, operated with a size parameter $R$ of 0.5.
The selection variables for the jets use the two highest $\pt$
jets in the event with pseudorapidity $|\eta|<$ 4.7.
The selection requirements are optimised
to obtain the best expected limit at 95\% CL on the VBF signal cross section
with fully simulated VBF signal events and the diphoton background
estimation from data~\cite{Hgamgam}.
The $\pt$ thresholds for the two jets are 30 and 20\GeV, and the pseudorapidity
separation between them is required to be greater than 3.5.
The dijet mass is required to be greater than 350\GeV.
Two selection criteria, relating the dijet to the diphoton
system, are applied: the difference
between the average pseudorapidity of the two jets and the pseudorapidity
of the diphoton system is required to be less than 2.5~\cite{Rainwater:1996ud},
and the difference in azimuthal angle between the diphoton system
and the dijet system is required to be greater than 2.6 radians.

\subsubsection{Lepton tag event classes}
Candidate diphoton events for the lepton-tagged channel have the same
selection requirements imposed on the photons as in the SM search~\cite{Hgamgam} except for
the $\pt$ thresholds.
As it is the case in the VBF, the \pt of the Higgs boson is also boosted in the VH production.
To maximize the signal efficiency, the photon \pt thresholds are set to
$\pt^{\gamma}(1)>45\times\mgg/120$, and $\pt^{\gamma}(2)>25\GeV$.

The lepton tag requires at least one muon or electron
with \pt $>$ 20\GeV, within $|\eta|<$ 2.4 for muons, and  $|\eta|<$ 2.5
for electrons.
Electrons are identified as a primary charged particle track and one or more ECAL energy clusters corresponding to this track extrapolation to the ECAL and to possible bremsstrahlung photons emitted along the way through the tracker material. Muons are identified as a track in the central tracker consistent with either a track or several hits in the muon system, not associated with a significant energy deposit in the calorimeters.

The leptons are required to be isolated, using isolation criteria
similar to those used for photons~\cite{Hgamgam}, and to be separated from the photons by
$\Delta R > 1$, where $\Delta R = \sqrt{\Delta\eta^2 + \Delta\phi^2}$.
To protect against the background events that arise from an electron
misidentified as a photon in the SM process $\cPZ\to\Pe\Pe$,
it is required that the mass of the photon-electron system
is not within $\pm$5\GeV of the nominal Z boson mass.

\subsubsection{Untagged event classes}
\label{sec:untag}
A substantial fraction of the H $\to\gamgam$ signal events are not
expected to pass either the dijet or lepton tag.
A statistically independent search is performed on untagged events by using
diphoton events that pass the \gamgam
selection, and photon \pt requirements of  $\pt^{\gamma}(1) > \mgg/3$
and $\pt^{\gamma}(2) > \mgg/4$,
but do not pass the selection for either of the two tagged channels.
Higgs bosons produced by VBF and VH mechanism have a harder transverse momentum spectrum than those
of the photon pairs produced by the background
processes~\cite{Ballestrero:2008gf}, and thus the background can be rejected while
retaining high signal efficiency, by placing a requirement on the
transverse momentum of the diphoton pair, $\ptgg$.
It is required that \piTgg$\equiv\ptgg/\mgg>0.1$.

The selected events are divided into four classes
according to the expected mass resolution and amount of background contamination~\cite{Hgamgam}.
Two photon classifiers are used: the minimum $\RNINE$ of the two photons,
$\RNINE^\text{min}$, and the maximum absolute pseudorapidity of the two photons.
The class boundary values for $\RNINE$ and pseudorapidity are chosen
to match those used to categorise photon candidates for photon identification cuts.
The untagged diphoton event classes are: (a) both photons in barrel
and $\RNINE^\text{min} >$ 0.94, (b) both photons in barrel and
$\RNINE^\text{min} <$ 0.94, (c) one or both photons in endcap and
$\RNINE^\text{min} >$ 0.94, and (d) one or both photons in endcap and
$\RNINE^\text{min} <$ 0.94.

\begin{table}[htbp]
\begin{center}
\topcaption{Number of selected events in the \gamgam event classes,
for data in the mass range $100-180\GeV$ and for an FP Higgs boson signal ($\mH$ = 120\GeV). The expected
number of background events in the signal region 115--125\GeV obtained from the background fit
and the mass resolution for the 120\GeV FP Higgs boson signal in each event class are also given.}
\begin{tabular}{|l|c|c|c|c|c|c|}
\hline
\multirow{3}{*}{} & Dijet & Lepton & \multicolumn{4}{c|}{Untagged} \\
\cline{4-7}
& tag  & tag & (a) & (b) & (c) & (d) \\
\hline
Data (100 $<\mgg<$ 180\GeV)  & 122 & 9 & 3866 & 5496 & 3043 & 4201 \\
Signal ($\mH = 120\GeV$) & 21.8 & 4.4 & 23.1 & 23.9 & 10.1 & 11.5 \\
Expected bkg  ($115 <\mgg< 125\GeV$) & 21.2 & 1.3 & 678.5 & 985.2 & 537.5 & 754.3 \\
\hline
$\sigma_\text{eff}$ (\GeVns) & 1.67 & 1.63 & 1.19 & 1.70 & 2.54 & 2.94 \\
\hline
\end{tabular}
\label{tab:ClassFracs}
\end{center}
\end{table}

The numbers of events in the \gamgam sub-channels are shown in
Table~\ref{tab:ClassFracs}, for simulated signal events and for data.
A Higgs boson with $\mH$ = 120\GeV is chosen for the signal, and the
data are counted in the mass range 100-180\GeV.
The table also shows the mass resolution, $\sigma_\text{eff}$,
defined as half the width of the narrowest window containing 68.3\% of the
distribution.

\subsubsection{Signal and background modelling}
\label{sec:ana}

In the SM $\PH\to\gamgam$ analysis~\cite{Hgamgam},
the diphoton mass spectrum of the signal and background are assumed to be described by analytical functions.
The signal shape was determined from the Z to electron mass spectrum,
while the background was described by smoothly falling analytical functions of various forms.
In this analysis an additional observable based on the diphoton
transverse momentum, \piTgg, is used to construct two-dimensional (2D) probability distribution function (PDF)
for the four untagged event classes.
This enables further exploitation of the difference between the signal
and background diphoton transverse momentum spectra.
Z boson decays to electrons are used to derive the amount of
additional smearing that needs to be applied to photons in MC simulated events
to reproduce the energy resolution observed in the data.
These smearing corrections are between 0.7 and 3\%, derived for photons separated into four $\eta$ regions
(two in the barrel and two in the endcap) and two categories of $\RNINE$.
The uncertainties on these corrections
and factors accounting for the difference between
photons and electrons are taken as  systematic
errors in the limit setting procedure.

The signal mass PDF, $\mathcal{M}_{s}(\mgg)$, is extracted, after
the smearing, by parameterising the $\mgg$ distribution in
simulated signal events with a sum of
Crystal Ball~\cite{CB} and Gaussian functions.

In the untagged \gamgam event classes, where a 2D analysis using \mgg
and \piTgg is performed, it is necessary to define a signal model that
is a function of these two observables, in the regions
$100< \mgg <180\GeV$ and $0.1< \piTgg <2$.
The correlations between the two variables are neglected,
because the mass resolution of the Higgs boson has little dependence on its momentum.
The PDF becomes a product of two one-dimensional PDFs,
one for the mass and one for the second observable, $\mathcal{K}_{s}(\piTgg)$,
empirically derived to be a sum of Gaussian and bifurcated Gaussian (a Gaussian PDF with different widths on left and right side of maximum value) functions.
The \piT shape uncertainty contributes less than 1\% to the expected exclusion limit.

\subsubsection{Background modelling in dijet and lepton tag classes}

For the dijet-tag event class, the background model is derived from data, by fitting
the diphoton mass distributions over the range $100< \mgg <180\GeV$.
The choice of the function used to fit the background and the choice of
the range are made based on a study of the possible bias introduced by the two choices.
The bias is studied for both the limit in the case of no signal and the measured signal strength in the case of a signal.

In case of the lepton-tag class, the requirement of an additional isolated lepton suppresses the
contribution from QCD background processes.
The remaining background is small, coming predominantly from
electroweak processes.
Its shape is derived from fitting the MC simulation for muon tags, and from fitting a combination of data and MC simulated events for electron tags.
For the electron-tagged events, a control sample (CS) is derived from data
by requiring one of the photons to be matched with a track.
This CS represents
the reducible background with enhanced statistics.
The final shape for this channel is the sum of the fits to simulation and the
CS, with the two components weighted by the cross sections of the
main irreducible and reducible background processes.
The sum is normalised to the data yield in the range 100--160\GeV.

Bias studies are performed using a number of generated pseudo-experiments with background only
and signal plus background hypotheses.
It is observed that using a second order polynomial fit
to the range $100< \mgg <180\GeV$ for the dijet-tagged events
results in only a small bias in either excluding or finding a
Higgs-boson signal in the mass range $110< \mgg <150\GeV$.
In both cases the maximum bias is found to be at
least five times smaller than the statistical uncertainties of the fit.
For the lepton-tagged classes, it is found, from the same technique, that the small number of selected events
allow the use of exponential functions for the fits without introducing
any significant bias as compared to the statistical uncertainty.

\begin{figure}[htbp]
   \begin{center}
      \includegraphics[width=0.80\linewidth]{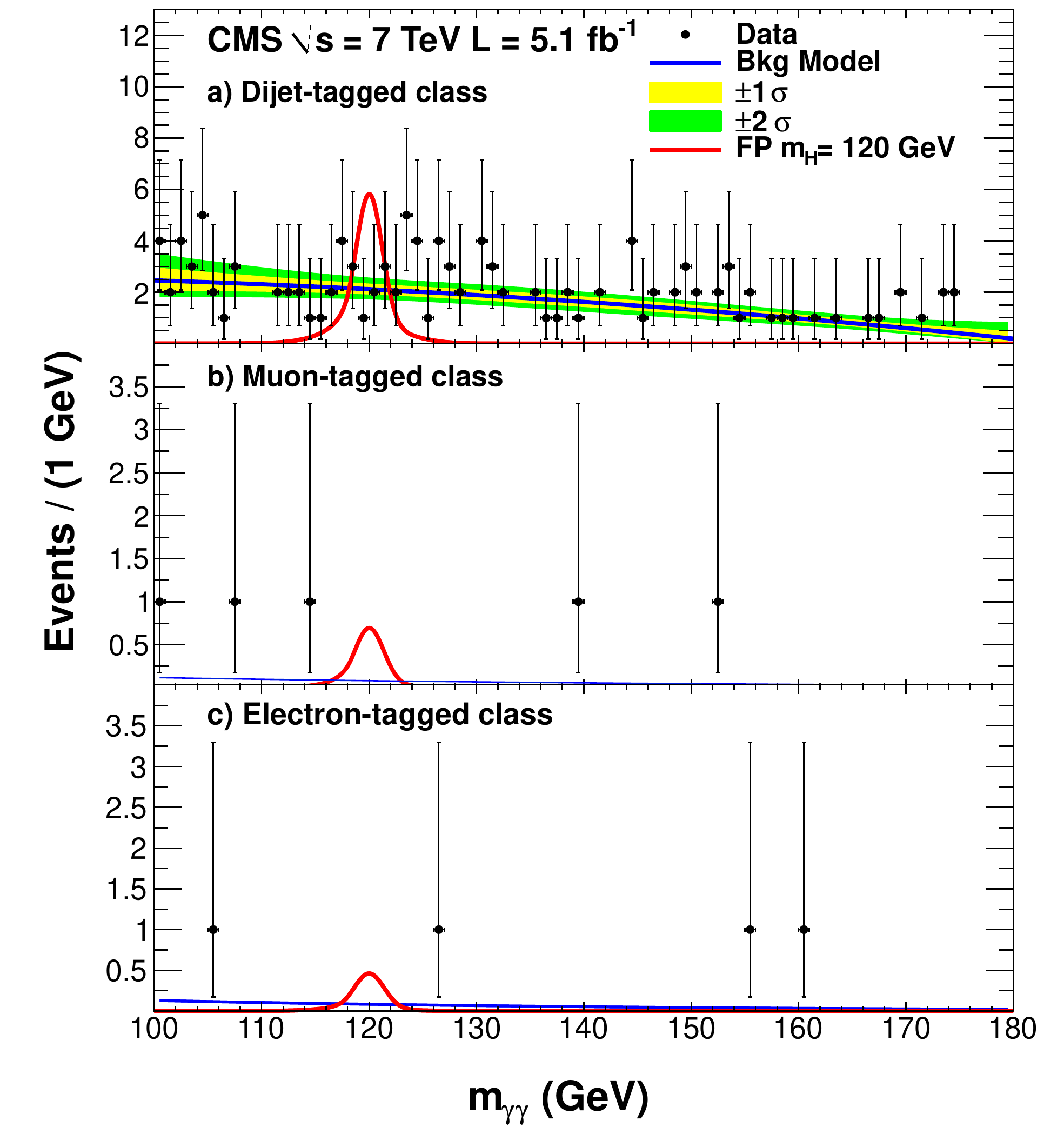}
     \caption{\label{fig:BckSigExcl}Data and the background model fits to the \mgg
distribution for the three diphoton event classes from tagging channels:
for dijet-tag with a second order polynomial fit (top), muon-tag (middle) and electron-tag (bottom) with an exponential fit for the lepton-tag. For the two bottom plots the uncertainty bands are not shown, as explained in the text. Signal for an FP Higgs with a mass of 120\GeV is overlaid for reference.}
   \end{center}
\end{figure}

The data distribution of \mgg and the corresponding background models
in the three tagged event classes are shown in Fig.~\ref{fig:BckSigExcl}.
For the dijet-tagged class the statistical uncertainty bands computed
from the fit are shown.
For both the muon-tagged and electron-tagged classes, uncertainty bands are not shown.
For these two classes the dominant statistical uncertainty on the background model is obtained from the number
of events which are used to determine the normalisation of the fit.

\subsubsection{Background modelling in the untagged classes}
\label{sec:InclModelling}

In the untagged \gamgam event classes, where a 2D analysis using \mgg
and \piTgg is performed, the background model
is a distribution of these two observables.
The nominal background PDF accounts for a linear correlation
between the two observables and has the following form:

\begin{equation}
\mathcal{P}_b = \mathcal{M}_b(\mgg, \piTgg | a_0, a_1)\times\mathcal{K}_b(\piTgg) = \mgg^{a_0+a_1\piTgg}\times\mathcal{K}_b(\piTgg)
\label{math:PDF-prod-corr}
\end{equation}

The empirical background PDF for the second observable, $\mathcal{K}_{b}(\piTgg)$ is defined as a sum
of an exponential function ($\mathcal{E}$) of slope $\tau_{B}$ and a Gaussian function ($\mathcal{G}$) of width $\sigma_{G}$ and mean fixed at zero:

\begin{equation}
\mathcal{K}_{b}(\piTgg|\tau_{B}, f_{d}, \sigma_{G} )= f_{d}\mathcal{E}(\piTgg| \tau_{B})+
(1-f_{d})\mathcal{G}(\piTgg| 0, \sigma_{G})
\end{equation}

\begin{figure*} [htbp]
\centering
\includegraphics[width=0.90\textwidth]{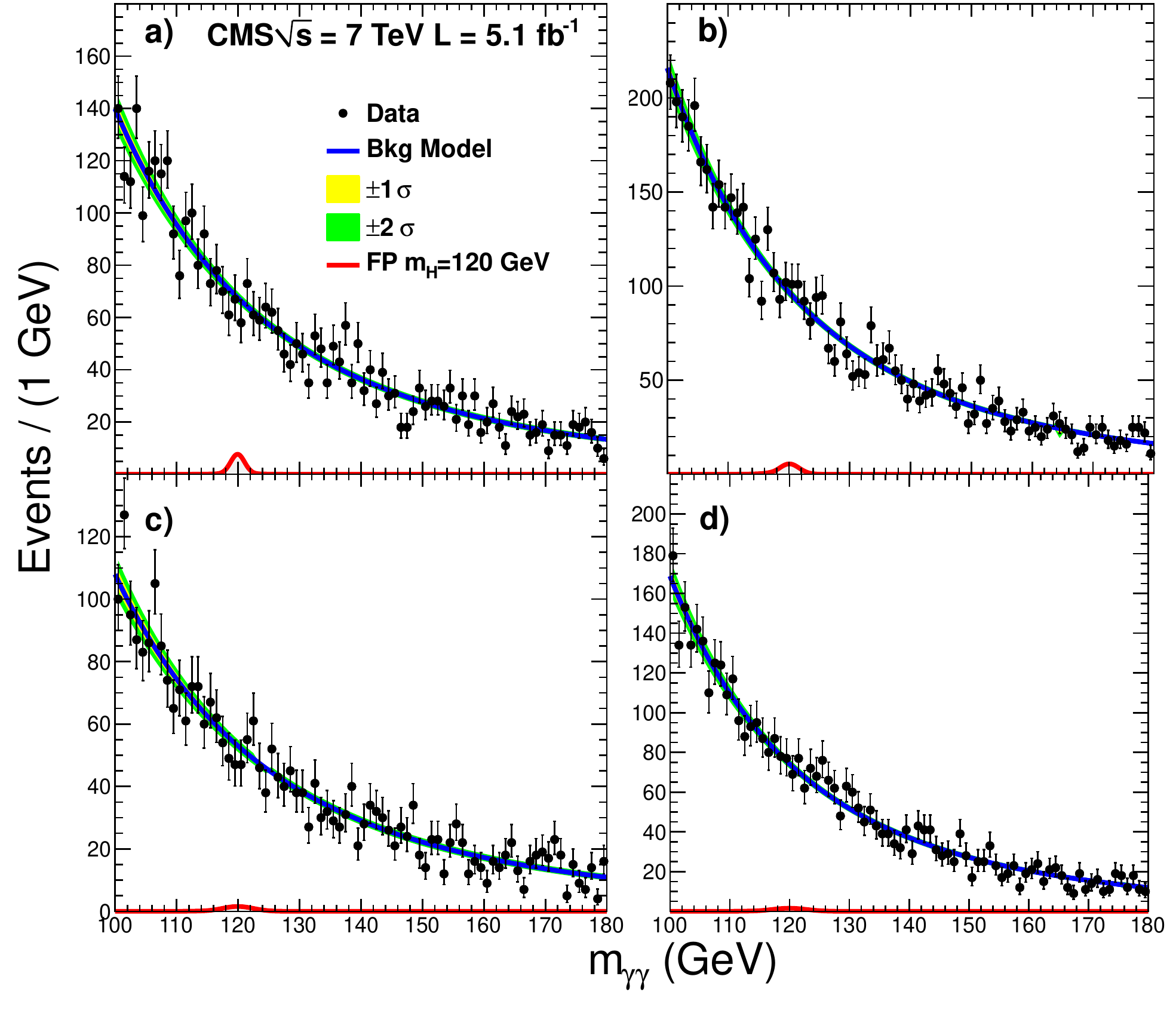}
\caption{
    {
    \mgg distribution in data (points) in the four categories of the
      untagged \gamgam sub-channel (a)--(d) defined in Section~\ref{sec:untag},
      together with background model fits of a power function including linear correlation to \piTgg.
      An MC simulated FP Higgs boson signal (\mH = 120\GeV) is overlaid for reference.}
    }
\label{fig:BckSigInclMass}
\end{figure*}

\begin{figure*} [htbp]
\centering
\includegraphics[width=0.90\textwidth]{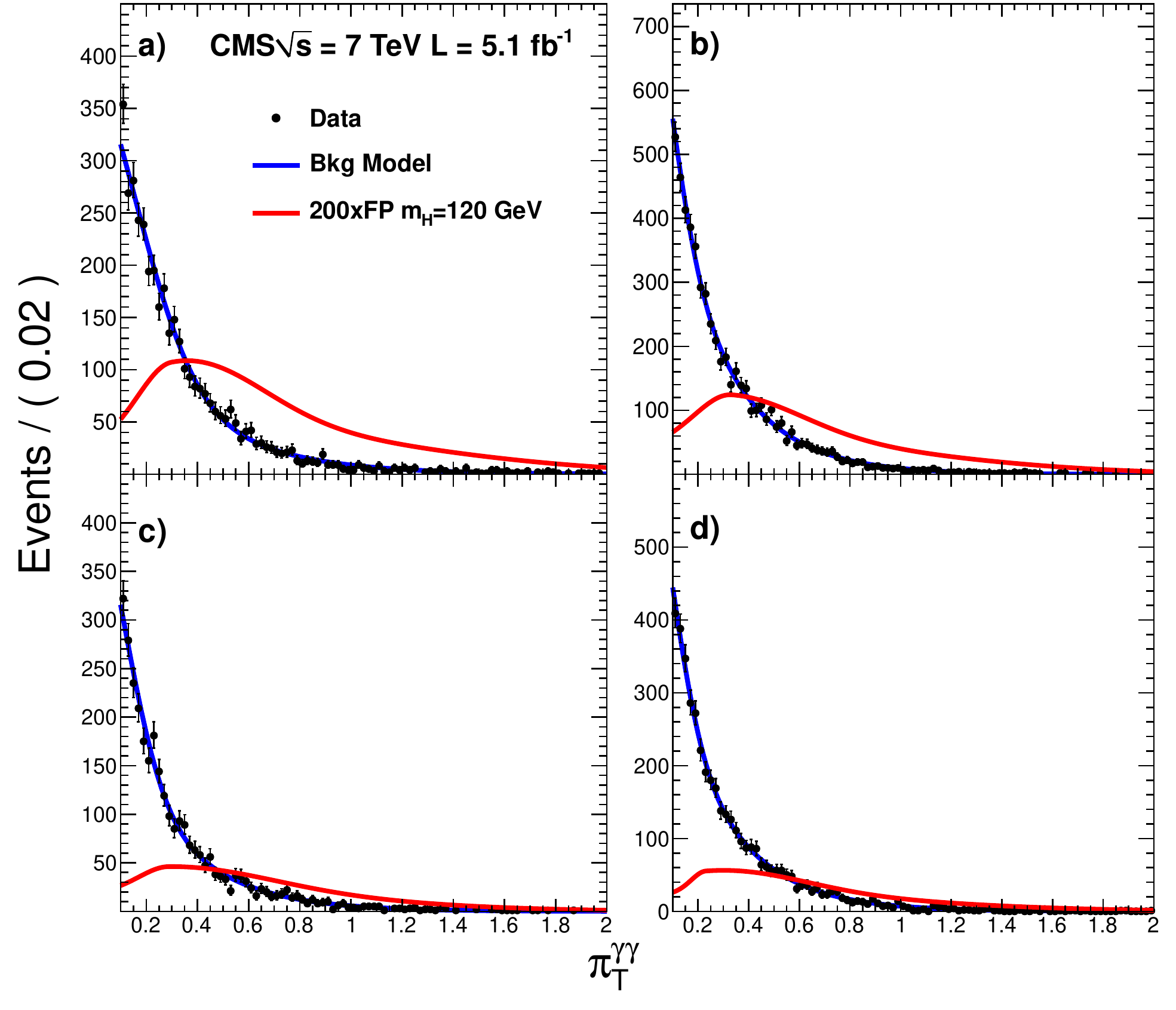}
\caption{
    {
    \piTgg distribution in data (points) in the four categories of the
    untagged \gamgam sub-channel (a)-(d) defined in Section~\ref{sec:untag}, together with
    background model fits of a sum of exponential and Gaussian functions
    centred at zero.
    A MC simulated FP Higgs boson signal (\mH = 120\GeV)  is also
    shown, scaled by 200 and modelled with a sum of a Gaussian and a
    bifurcated Gaussian.}
    }
\label{fig:BckSigInclpiT}
\end{figure*}

A power law function is chosen to describe the \mgg distribution.
The data in each of the four untagged event classes are fitted separately.
Figures~\ref{fig:BckSigInclMass} and~\ref{fig:BckSigInclpiT} show
the data and the fit results projected on \mgg, and on \piTgg, respectively, for each
class.

Goodness-of-fit tests are performed measuring the bias of the model due to correlations
between the two observables and due to choice of functional forms.
Pseudo-experiments are performed, generated from alternative background models,
and the signal plus background model is fitted for various test masses.
The bias is taken as the mean of the pull distribution, which is defined as the difference between
the fitted and generated signal strength divided by the
statistical error from the fit in each event.
If the pseudo-experiments are generated from a background model
containing a linear correlation of the two variables, a maximum bias of 60\% is observed if a linear correlation is not
included in the fitting function.
Tests have shown that with a linear plus quadratic correlation in the model
a fit with only a linear correlation results in a bias of less than 13\% in the entire fit range.
This bias is regarded as negligible and thus it is concluded that a
fit function with linear correlation is adequate.
With a similar procedure, the nominal fit function is tested against alternative models of mass shape
with functions including linear correlations.
A maximum bias of about one quarter
of the statistical error is measured, which is negligible.

\subsection{Diboson WW decay mode}

\subsubsection{Dijet tag event class}

\begin{figure}[htbp]
   \begin{center}
      \includegraphics[width=0.48\linewidth]{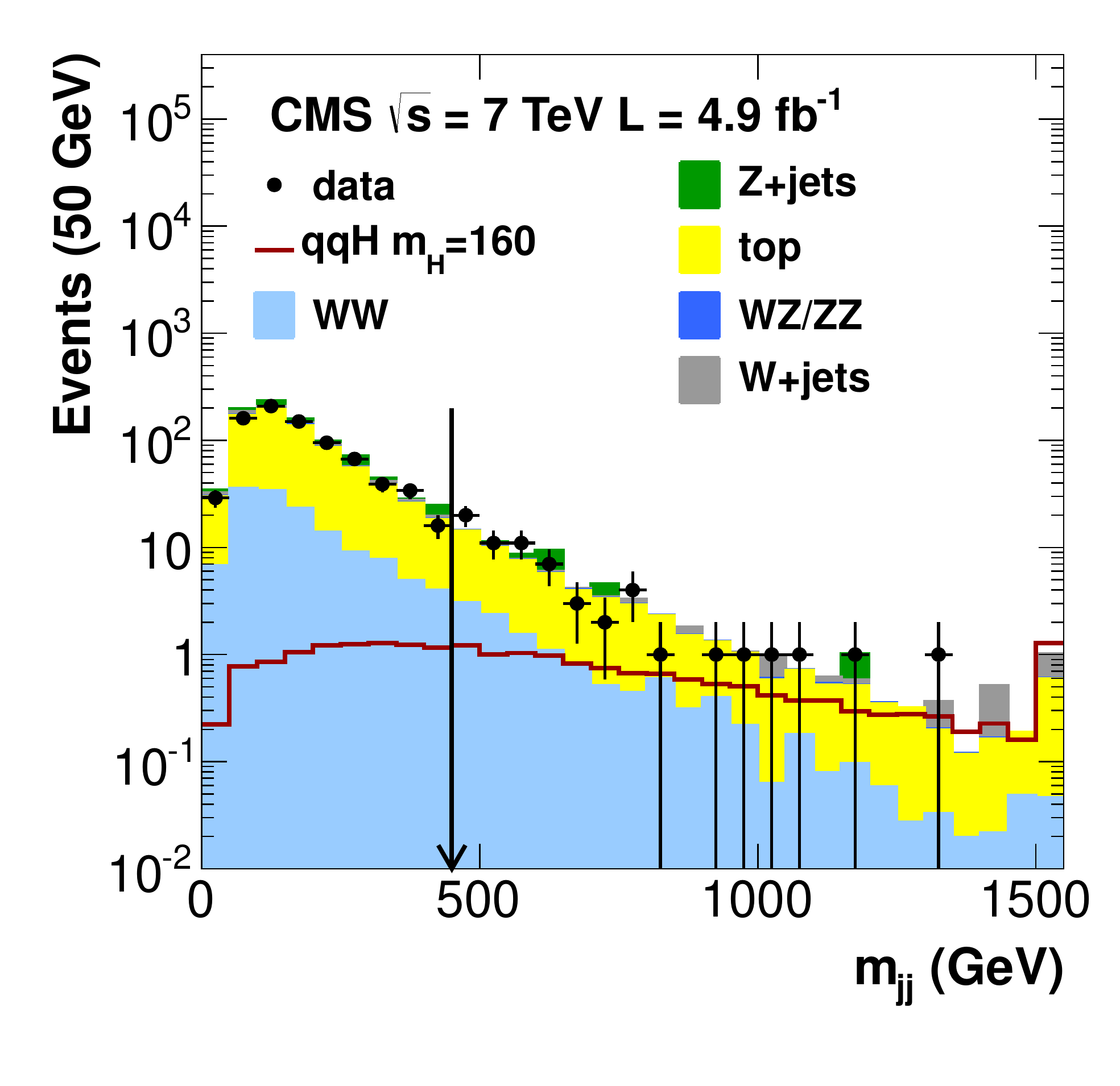}
      \includegraphics[width=0.48\linewidth]{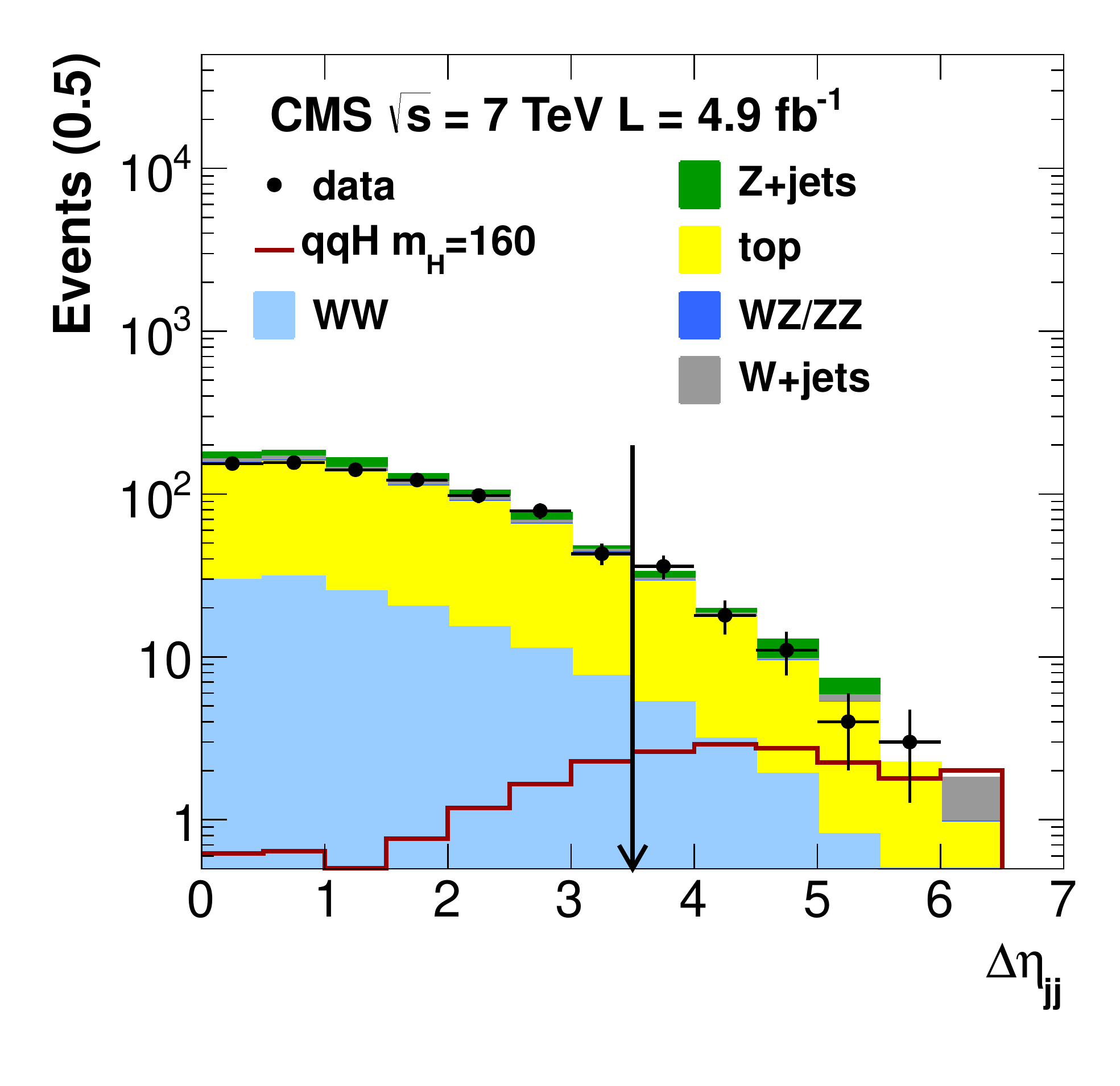}
     \caption{The distributions of $m_{j_1j_2}$ (left) and
       $\Delta\eta_{jj} = |\eta_{j_1}-\eta_{j_2}|$ (right)
              after the WW selection.
              Contributions from backgrounds are shown together with data,
              and the expected signal from VBF production
              is shown for $\mH = 160\GeV$.
							The arrows show the minimum values in the selections applied for the WW dijet-tag analysis.
             }
     \label{fig:WW_VBF_plots}
   \end{center}
\end{figure}

The $\PH \to \PW\PW^{(*)} \to 2\ell 2\cPgn$ analysis~\cite{HWW}
selects events with two isolated leptons of opposite charge,
large missing transverse energy, and two jets with VBF topology.
One lepton is required to have $\pt>20\GeV$, while the second is required to have
$\pt>10\GeV$.
The fiducial region is $|\eta|<2.4$ for muons and $|\eta|<2.5$ for electrons.
If the two leptons have the same flavour,
the one with lower \pt{} is required to have a $\pt$ of at least 15\GeV
to suppress the Drell-Yan background.
The missing transverse energy requirement is applied
by means of a selection on the projected $\met$
defined as the component of $\met$
transverse to the nearest lepton if that lepton is within $\pi$/2 in
azimuthal angle, or the full $\met$ otherwise~\cite{CDF},
which is required to be larger than about 40\GeV,
the precise value depending on the number of vertices found in each event.
Jets are required to have a $\pt > 30\GeV$ and $|\eta| < 5$,
and the two jets with the highest \pt are chosen as tag jets ($j_1$, $j_2$).
The VBF selection requires $|\eta_{j_1}-\eta_{j_2}| > 3.5$ and
$m_{j_1j_2} > 450\GeV$.
The distributions of $m_{j_1j_2}$ and $\Delta\eta_{jj} = |\eta_{j_1}-\eta_{j_2}|$
are shown in Fig.~\ref{fig:WW_VBF_plots}
after the WW selection.
Besides the cuts on the dijet system,
it is required that the event has no other jets above 30\GeV
found between the tag jets in pseudorapidity.
These selections suppress the dominant background coming from top-quark production,
which is also reduced by
requiring a b-veto on the tag jets using jet impact parameters.
The Drell-Yan background is suppressed by requirements on the dilepton system mass and momentum,
as well as on the angle between the dilepton and the dijet system.
Events with additional leptons above 10\GeV are rejected.
After all requirements, between 10 and 20 data events
remain in the signal regions which are defined according to the Higgs boson mass hypothesis~\cite{HWW}.
These events are compatible with a background only hypothesis.
The main backgrounds, namely $\ttbar$ and $\Zjets$,
are estimated from the data:
the $\ttbar$ is measured in the region where the central jet is b-tagged,
while the $\Zjets$ is measured under the Z peak.
Also the contamination coming from $\Wjets{}$ and QCD
is measured in data, in a phase space with relaxed lepton identification.
The WW background is evaluated from simulation.

\subsubsection{Zero and one-jet event classes}

The results obtained by the inclusive analysis~\cite{HWW} in the zero and one-jet classes
are also included in the limit extraction.
The electron and muon selection is the same as for the dijet-tagged class.
A \pt threshold of 30\GeV is placed on the jets,
the number of which defines the classes.
Events are split into same-flavour and different-flavour dilepton
sub-channels, since the background from Drell-Yan production is much
larger for the same-flavour dilepton events.
The dominant background for these classes is from WW,
together with \Zjets~and top-quark production in the one-jet class,
as well as contaminations from \Wjets{} and QCD.

\subsubsection{Lepton tag event class}

The $\PW \PH \to \PW\PW\PW \to 3\ell3\nu$ analysis
selects events with three charged leptons, either electrons or muons, large
$\ETmiss$, and low hadronic activity.
The third lepton has to be isolated and have $\pt >10\GeV$,
and it is required that there be no jet with $\pt >40\GeV$ in the event.
The dominant background comes from
$\PW\cPZ \to 3\ell\nu$ production, which is largely eliminated by
requiring that the invariant mass of all
same-flavour oppositely charged lepton pairs
is not within $\pm$25\GeV of the nominal Z boson mass.
In addition, the smallest dilepton mass $m_{\ell{}\ell{}}$ constructed from
oppositely charged leptons is required to lie between 12 and 100\GeV,
and the smallest distance, $\Delta R$, between them is required to be less than 2.
The background processes with jets misidentified as leptons, e.g. $\cPZ$+jets and top,
as well as the $\PW\cPZ \to 3\ell\nu$ background are estimated from data.
The small contribution from the $\cPZ\cPZ \to 4\ell$ process with one unreconstructed lepton
is estimated using simulated samples.
After all cuts, 7 data events remain in the signal region
while $8.4\pm0.9$ events are expected from simulation.

\subsubsection{Signal and background modelling}
\label{sec:VVmodels}

For the dijet and lepton tag WW classes,
the hypothesis testing is based on the number of expected signal and backgrounds events only.
Because of the impossibility to fully reconstruct the Higgs resonance and the small number of events expected
in these classes, the limit extraction is based on counting experiments in both the VBF and WH sub-channels.
The number of expected signal events is evaluated from simulation,
while the background contamination in the signal regions
is estimated with methods based on data whenever possible.
For the zero and one jet classes,
the limit extraction is based on the shape of a multi-variate discriminant,
optimised to maximise the difference between the signal and the WW background.
The discriminant is built on the kinematics of the dilepton pair and the missing energy.
For the signal case,
the model is obtained from the simulation;
for most of the backgrounds,
the templates are taken from the simulation and cross-checked in control samples in data.
For the \Wjets{} background
the nominal shape is derived from a data control sample.

\subsection{Diboson ZZ decay mode}
\label{sec:ZZclasses}

In the $\PH \to \cPZ\cPZ^{(*)} \to 4\ell$ channel~\cite{HZZ4l}
a search is made for a four-lepton mass peak over a small continuum background.
The $4\Pe$, $4\Pgm$, $2\Pe2\Pgm$ sub-channels are analysed separately
since there are differences in the four-lepton mass resolutions
and the background rates arising from jets misidentified as leptons.
The dominant irreducible background in this channel is from non-resonant $\cPZ\cPZ$ production
(with both $\cPZ$ bosons decaying to either $2\Pe$, or $2\Pgm$, or $2\Pgt$ with the taus decaying leptonically)
and is estimated from simulation.
The smaller reducible backgrounds with jets misidentified as leptons,
e.g. $\cPZ+\text{jets}$, are estimated from data.

In the $\PH\to\cPZ\cPZ  \to 2\ell 2\cPgn$ search~\cite{HZZ2l2nu},
events are selected by the presence of a lepton pair ($\Pe\Pe$ or $\Pgm\Pgm$),
with invariant mass consistent with that of an on-shell $\cPZ$ boson,
and a large $\ETmiss$.
A transverse invariant mass $m_{\mathrm{T}}$ is defined from the dilepton momenta
and $\ETmiss$, assuming that $\ETmiss$ arises from a $\cPZ \to \cPgn\cPgn$ decay.
A broad excess of events is searched for in the $m_{\mathrm{T}}$ distribution.
The non-resonant $\cPZ\cPZ$ and $\PW\cPZ$ backgrounds are taken from simulation,
while all other backgrounds
are evaluated from control samples in data.

In the $\PH \to \cPZ\cPZ^{(*)} \to 2\ell 2\cPq$ search~\cite{HZZ2l2q},
events are selected with two leptons ($\Pe\Pe$ or $\Pgm\Pgm$) and
two jets with zero, one, or two $\cPqb$-tags, thus defining a total of six exclusive final states.
Requiring $\cPqb$-tagging improves the signal-to-background ratio.
The two jets are required to form an invariant mass
consistent with that of an on-shell $\cPZ$ boson.
The aim is to search for a peak in the invariant mass distribution
of the dilepton-dijet system, with the background rate and shape estimated using control regions in data.

In the $\PH\to\cPZ\cPZ \to 2\ell 2\Pgt$ search~\cite{HZZ2l2tau},
one $\cPZ$ boson is required to be on-shell and to decay to a lepton pair ($\Pe\Pe$ or $\Pgm\Pgm$).
The other $\cPZ$ boson is required to decay through a $\Pgt\Pgt$ pair to one
of the four final-state signatures
$\Pe\Pgm$, $\Pe \Pgt_{\mathrm{h}}$, $\Pgm \Pgt_{\mathrm{h}}$, $\Pgt_{\mathrm{h}}\Pgt_{\mathrm{h}}$,
where $\Pgt_{\mathrm{h}}$ is a hadronically decaying $\Pgt$.
Thus, eight exclusive sub-channels are defined.
A broad excess is searched for
in the distribution of the dilepton-ditau mass,
constructed from the visible products of the tau decays,
neglecting the effect of the accompanying neutrinos.
The dominant background is non-resonant  $\cPZ\cPZ$ production whose rate is estimated from simulation.
The main sub-leading backgrounds with jets misidentified as $\tau$ leptons
stem from $\cPZ+\text{jets}$ (including $\cPZ\PW$) and top-quark events.
These backgrounds are estimated from data.

\subsubsection{Signal and background modelling}
\label{sec:VVmodelsTwo}

The limit calculation for the ZZ classes is based on the shape of the invariant mass distribution
of the decay products,
and the likelihood is written in terms of the estimated
probability distribution function
for the signal and the background.
For $\PH \to \cPZ\cPZ^{(*)} \to 4\ell$ and $\PH \to \cPZ\cPZ^{(*)} \to 2\ell 2\cPq$,
the signal shapes are described by means of analytical fits,
based on the Crystal Ball function.
For the $\PH\to\cPZ\cPZ \to 2\ell 2\Pgt$ class the likelihood is based on the binned
distribution of the reconstructed visible mass.
For the $\PH\to\cPZ\cPZ \to 2\ell 2\cPgn$ class the likelihood is based on the binned
distribution of the transverse mass calculated using the visible decay products.
The background shapes are extracted from data when possible,
while for the irreducible ones, such as the electroweak ZZ or WZ production,
the simulation is used.

\section{Results}
\label{sec:res}
The statistical approach considered in evaluating the limit is the asymptotic \CLS~\cite{Asympt}
with the profile likelihood ratio as a test-statistic~\cite{LHC-HCG}. %\cite{CMScombFeb2012}. %PG FIXME this is a PAS
Given the narrowness of the Higgs mass peak in the \gamgam channel, which has a resolution approaching 1\GeV in the classes
with the best resolution, the search is carried out with steps of 0.5\GeV in the signal hypothesis mass in the range between 110 and 150\GeV.
All known sources of systematic uncertainties are included in the likelihood model which is used for the limit setting.
Systematic errors which are correlated between event classes (theory, luminosity, photon and trigger efficiency, etc)
are included as common nuisance parameters.

Following the prescription in~\cite{Dittmaier:2012vm},
the QCD scale uncertainties on the FP Higgs boson production cross
section are increased with respect to those of the SM Higgs boson, to 5\%,
to cover the effects of electroweak corrections which have not yet been calculated.

\begin{figure}[hbtp]
\begin{center}
      \includegraphics[width=0.48\textwidth, angle=0]{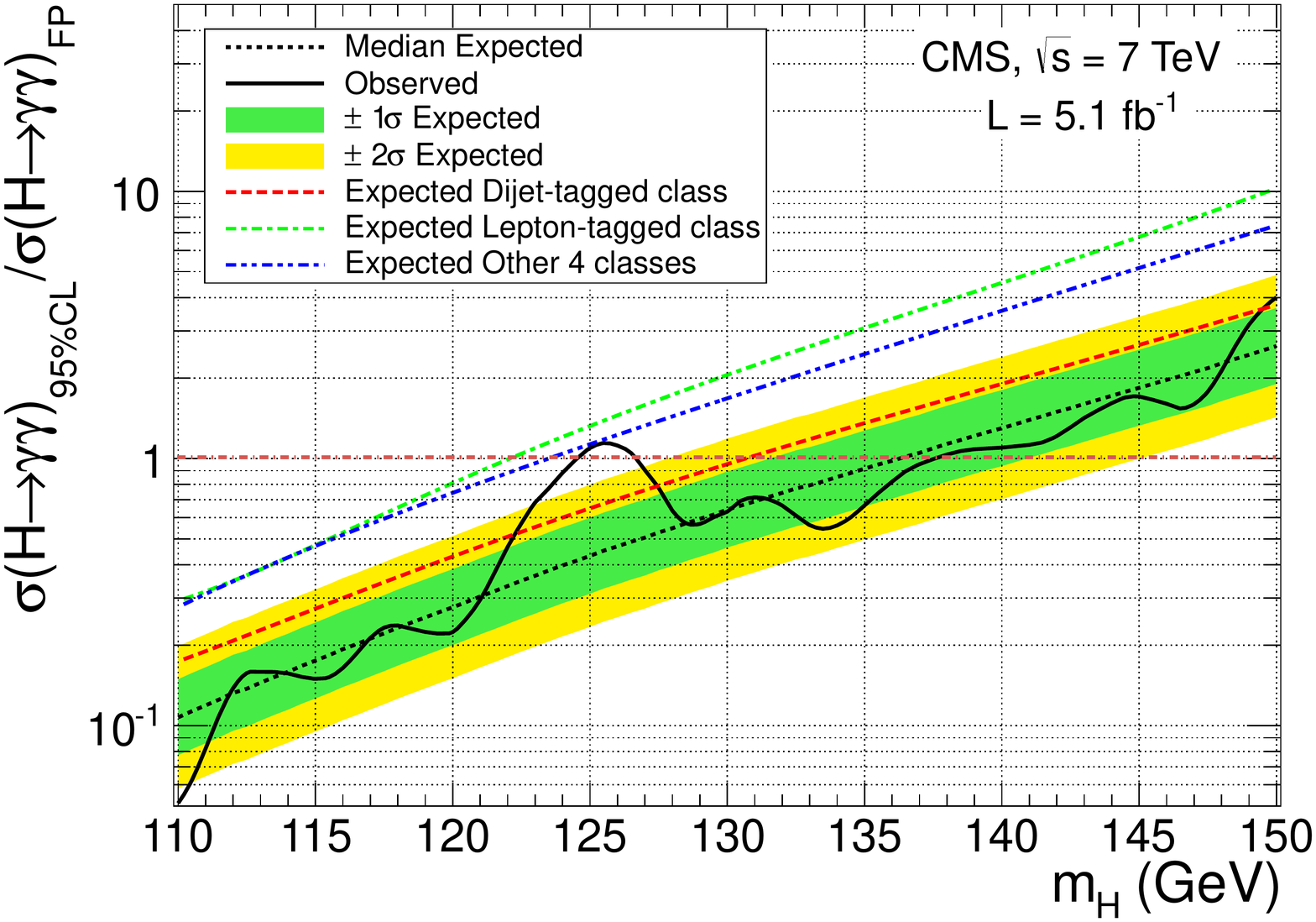}
      \includegraphics[width=0.48\linewidth]{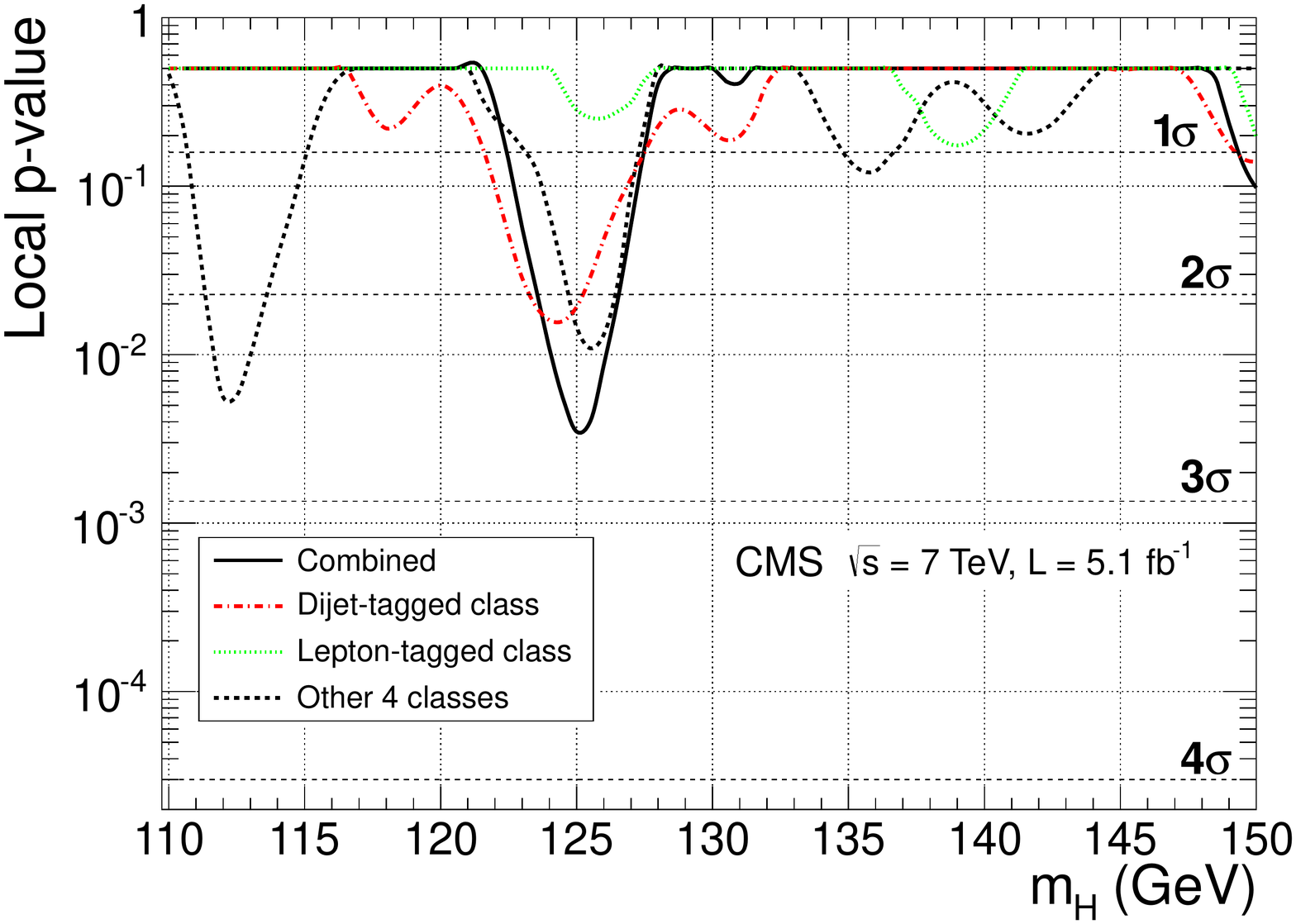} \\
        \caption{(Left) The 95\% CL upper limit on the signal
          strength, $\sigma/\sigma_{FP}$, of an FP Higgs boson, as a
          function of the Higgs boson mass, for the \gamgam channel.
         The dashed line indicates the expected median of results for the background-only hypothesis,
         while the two bands indicate the ranges that are expected to contain
         68\% and 95\% of all observed excursions from the median, respectively.
        The asymptotic \CLS method is used. Individual contributions to the expected limit for each of the channels are shown with dotted lines.
	(Right) The $p$-value of \gamgam channel. Contributions of individual sub-channels are also shown.
        \label{fig:GamGamLimit}
             }
\end{center}
\end{figure}

Figure~\ref{fig:GamGamLimit} (left) shows the limit relative to the FP
model expectation from the \gamgam sub-channels only, where the systematic uncertainties
on the expected cross section and branching fraction are included in the limit setting procedure.
The observed values are shown by a solid line.
The contributions to the expected limit of each of the \gamgam
sub-channels are also shown.
The sensitivity of the search in this channel lies predominantly in the dijet tag sub-channel.
The \gamgam combined expected exclusion limit at 95\% CL covers the mass range between 110--136.5\GeV, while the data
exclude ranges from 110--124.5\GeV and 127--137.5\GeV.
Figure~\ref{fig:GamGamLimit} (right) shows the local $p$-value for \gamgam channel and each sub-channel, calculated
from the asymptotic approximation~\cite{LHC-HCG}, at 0.5\GeV
intervals in the mass range 110--150\GeV.
The local $p$-value quantifies the probability for the background to
produce a fluctuation at least as large as observed, and
assumes that the relative signal strength between the
event classes follows the MC signal model for the FP Higgs boson.
The local $p$-value corresponding to the largest upwards fluctuation
of the observed limit, at \FPMaxZmass\GeV, has been computed to be
3.6$\times$10$^{-3}$ (2.7$\sigma$) in the asymptotic approximation.
When taking into consideration the look-elsewhere effect~\cite{LEE} in the search range 110$-$150\GeV,
the global significance of this deviation is 1.2~$\sigma$.

\begin{figure}[hbtp]
\begin{center}
       \includegraphics[width=0.48\textwidth]{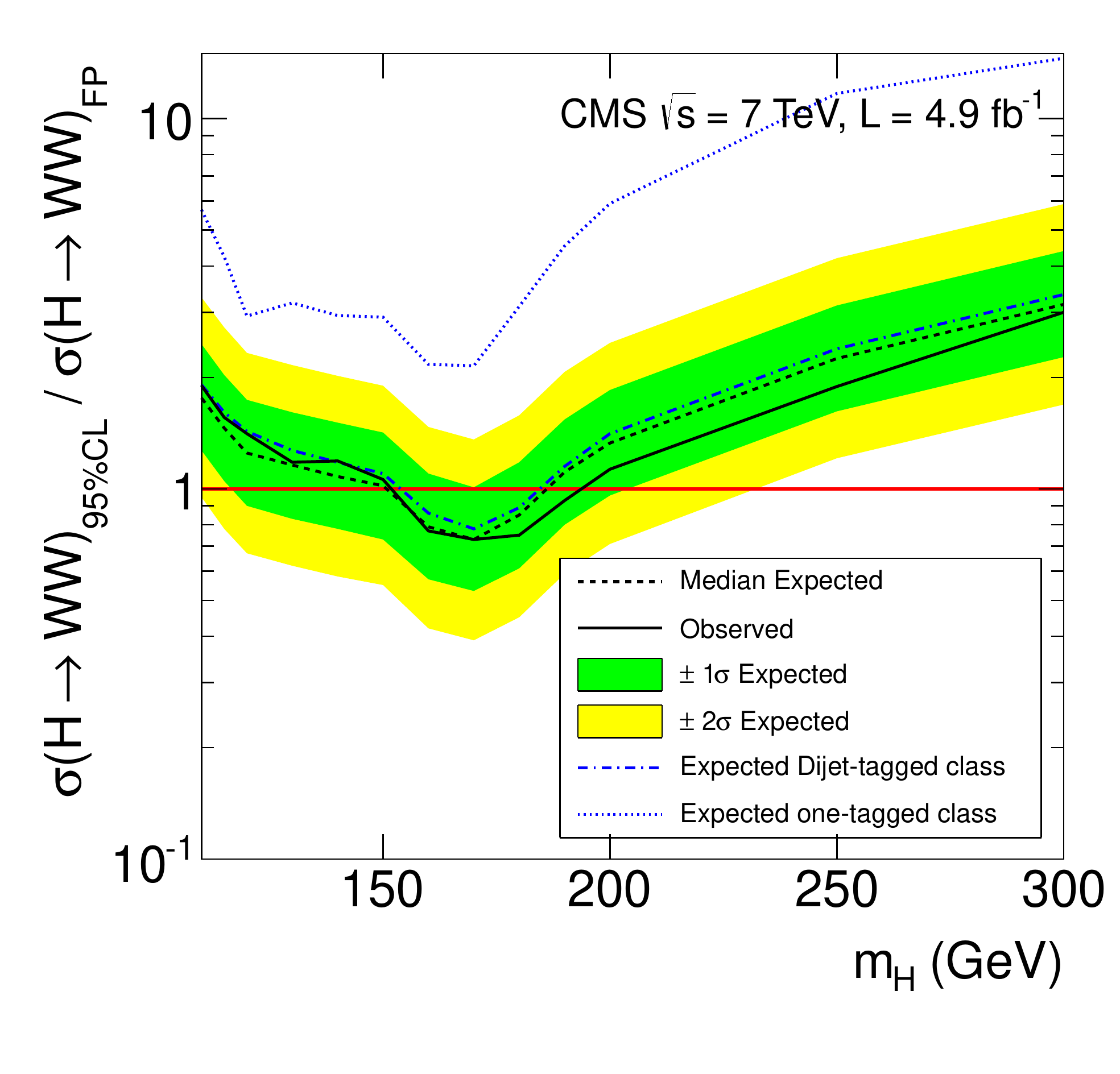}
      \includegraphics[width=0.48\linewidth]{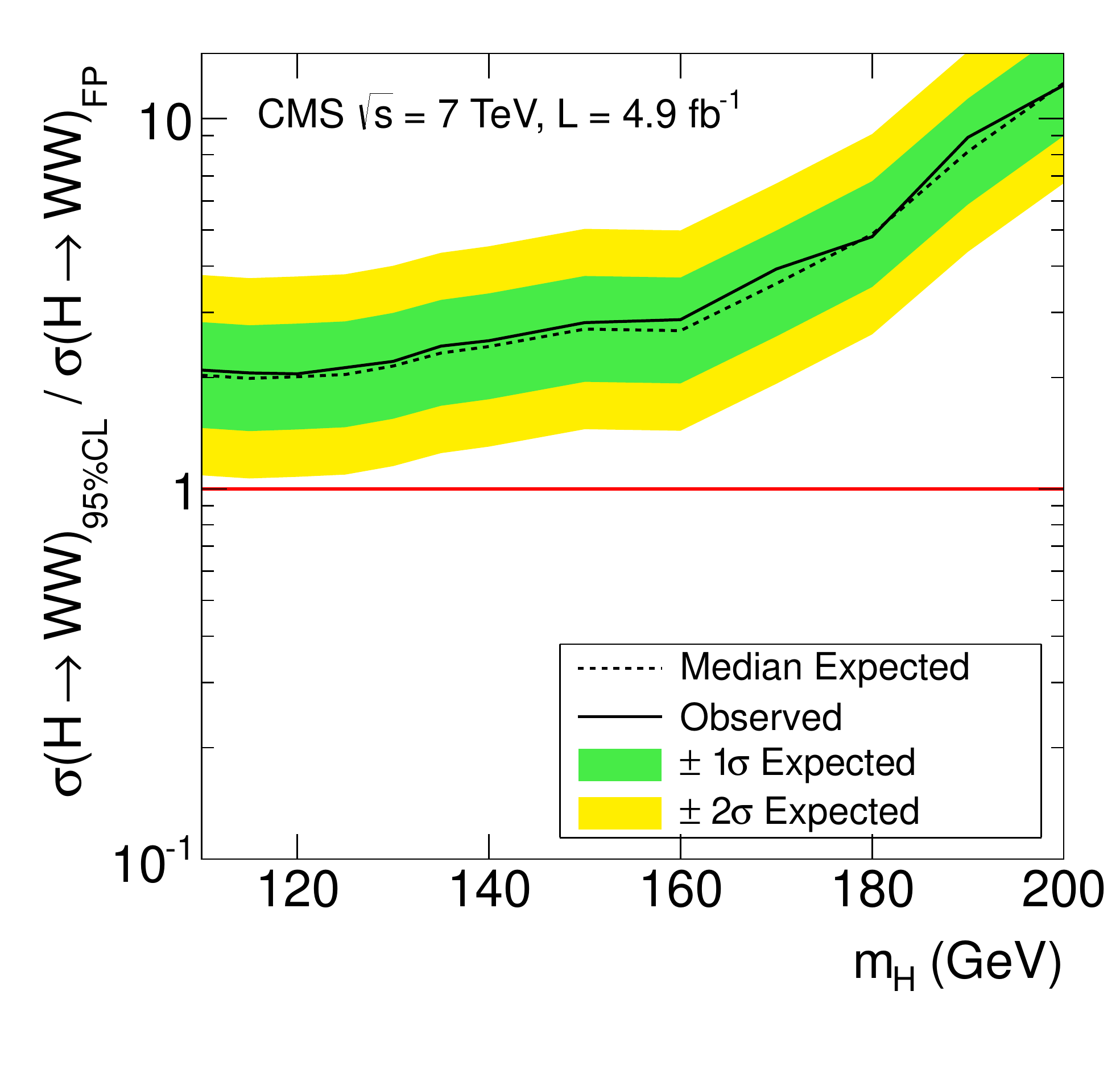}
        \caption{The 95\% CL upper limit on the signal
          strength, $\sigma/\sigma_{FP}$, of an FP Higgs boson, as a
          function of the Higgs boson mass, for the WW channel.
        (Left) Limit for the WW final state excluding the WH trilepton
        sub-channel. The contributions to the expected limit from the
        dijet-tagged and the single jet event classes are also shown.
        (Right) Limit for the WH trilepton sub-channel.
         The dashed line indicates the expected median of results for the background-only hypothesis,
         while the two bands indicate the ranges that are expected to contain
         68\% and 95\% of all observed excursions from the median, respectively.
        The asymptotic \CLS method is used.
        \label{fig:WWLimit}
             }
\end{center}
\end{figure}

Figure~\ref{fig:WWLimit} shows the 95\% CL upper limit on the signal
strength, $\sigma/\sigma_{FP}$, of an FP Higgs boson, as a
function of the Higgs boson mass, for the WW channel.
The contributions from the individual sub-channels are indicated.
The limit from the dijet-tagged sub-channel complements the \gamgam search channels,
excluding the FP hypothesis from 146\GeV to 196\GeV.

The 32 sub-channels of the three decay modes, \gamgam, WW, and ZZ, described in Section 3,
are combined using the combination techniques described in Ref.~\cite{LHC-HCG}
to account for all statistical and systematic uncertainties and their correlations.
The uncertainties consist of:
theoretical uncertainties on the expected cross sections and acceptances for signal and background processes,
experimental uncertainties in the modeling of the detector response (event reconstruction and selection efficiencies, energy scale and resolution),
and statistical uncertainties associated with either ancillary measurements of
backgrounds in control regions or selection efficiencies obtained using simulated events.

\begin{figure*} [htbp]
\centering
\includegraphics[width=0.46\textwidth]{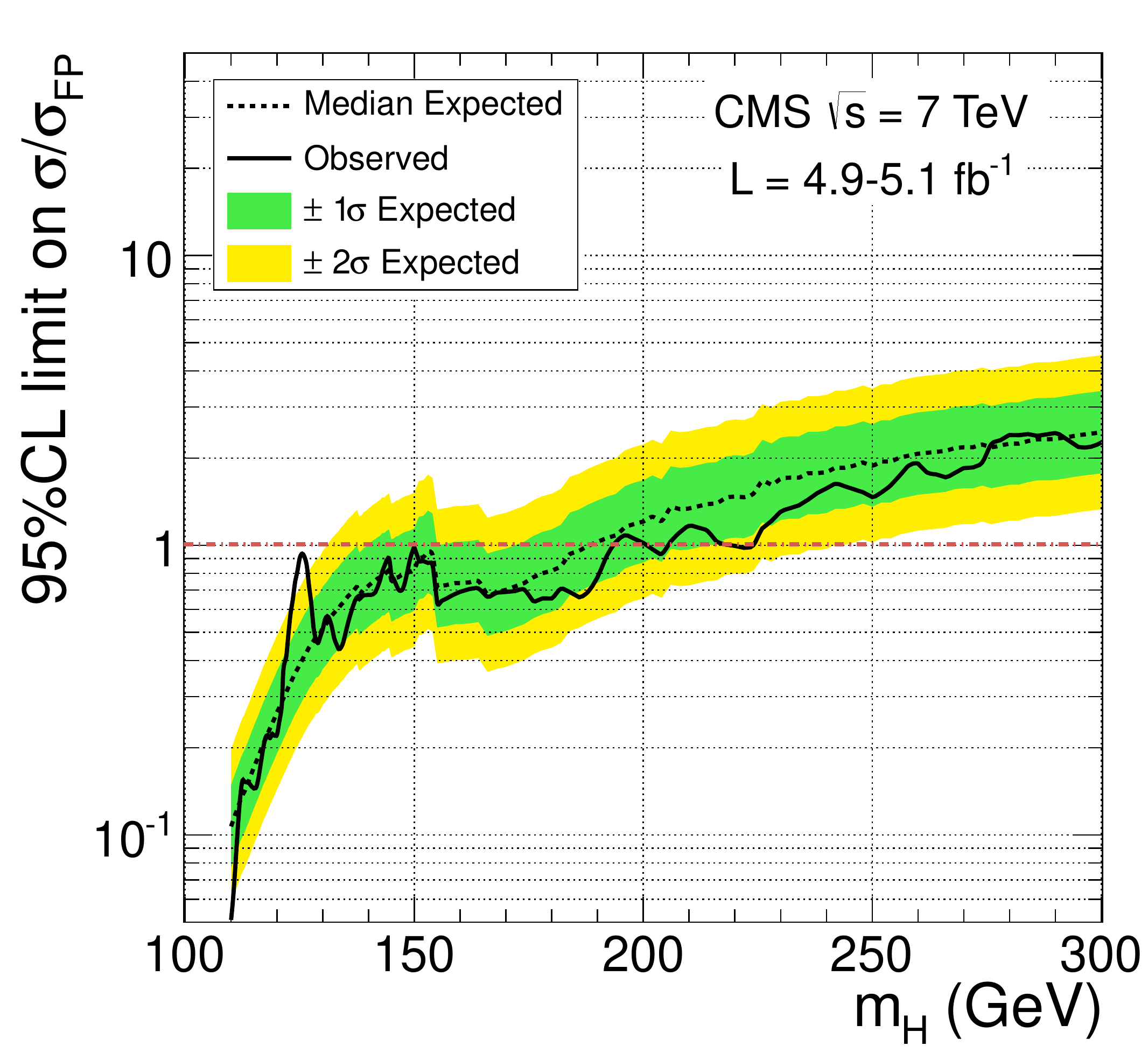} \hfill
\includegraphics[width=0.48\textwidth]{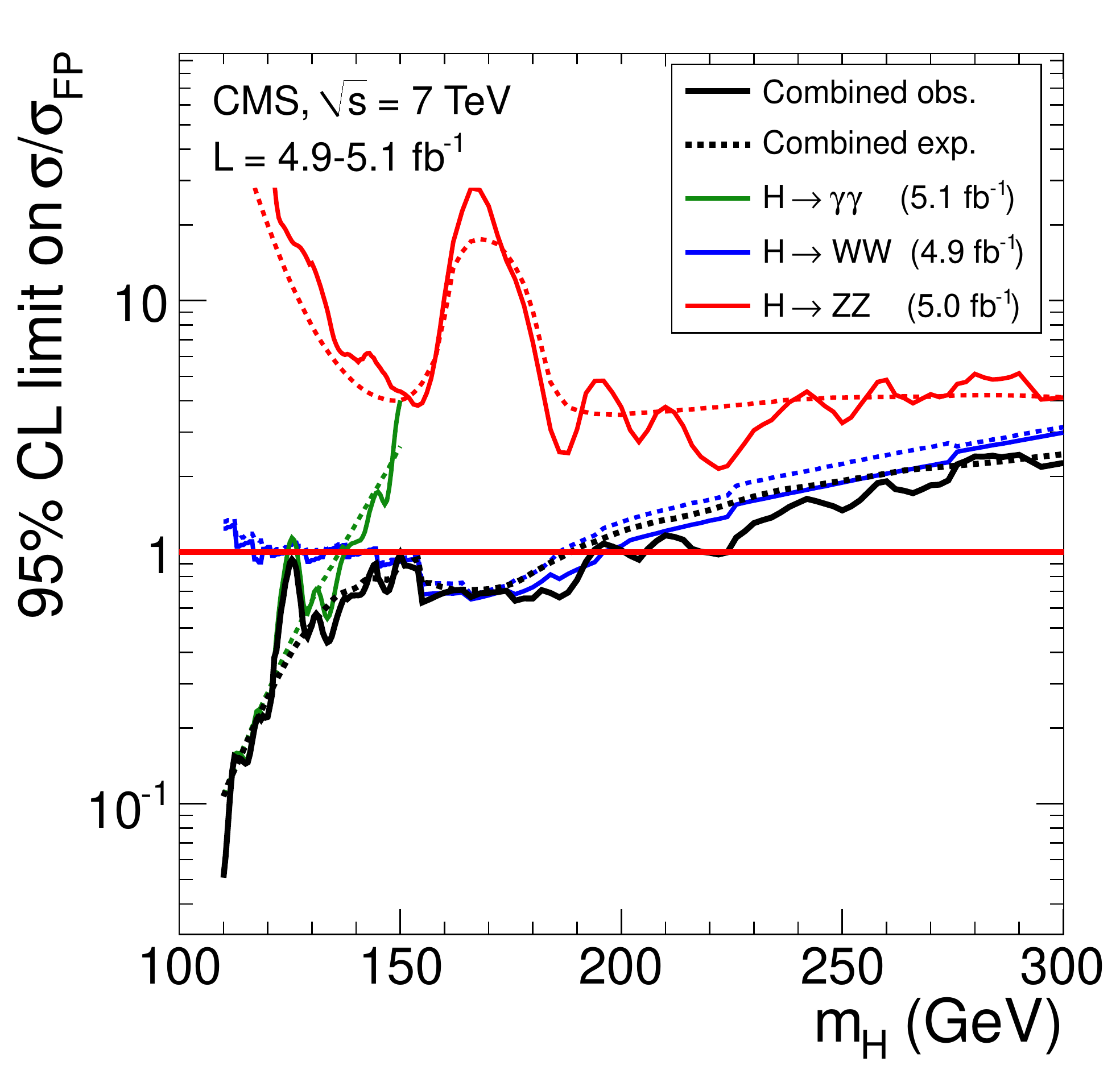}
\caption{
    {\color{black}
    (Left) The {\color{black} observed and expected} 95\% CL upper limits on the signal strength,
    $\sigma / \sigma_\text{FP}$,
    for the fermiophobic Higgs boson hypothesis as a function of the Higgs boson mass.
    (Right): The observed and expected 95\% CL upper limits on the signal strength
    as a function of the fermiophobic Higgs boson mass
    for the {\color{black} three explored} Higgs boson decay modes and their
    combination. }
    }
\label{fig:FP_Mu95}
\end{figure*}

The limit on the signal strength, $\sigma/\sigma_{FP}$, of an FP Higgs boson, as a
function of the Higgs boson mass,
calculated using the asymptotic approximation, is shown in
Fig.~\ref{fig:FP_Mu95}, together with the expected and observed 95\%~CL limits
for individual fermiophobic Higgs boson decay modes as well as for their combination.
Checks at a few test points around 125\GeV have shown
the calculation to be consistent with values obtained by the
full modified frequentist approach~\cite{Read1}.
The fermiophobic Higgs boson is excluded at 95\% CL in the mass range \FPObsA --\FPObsB\GeV.
At 99\%~CL, we exclude the fermiophobic Higgs boson in the range \FPObsAA --\FPObsBB\GeV,
with the exception of two gaps: \FPObsGapOneAA --\FPObsGapOneBB\GeV ~and \FPObsGapTwoAA--\FPObsGapTwoBB\GeV.
The sensitivity of the search lies predominantly in the \gamgam channel below 140\GeV, and in the WW channel for the high mass search range.

\begin{figure*} [t]
\centering
\includegraphics[width=0.46\textwidth]{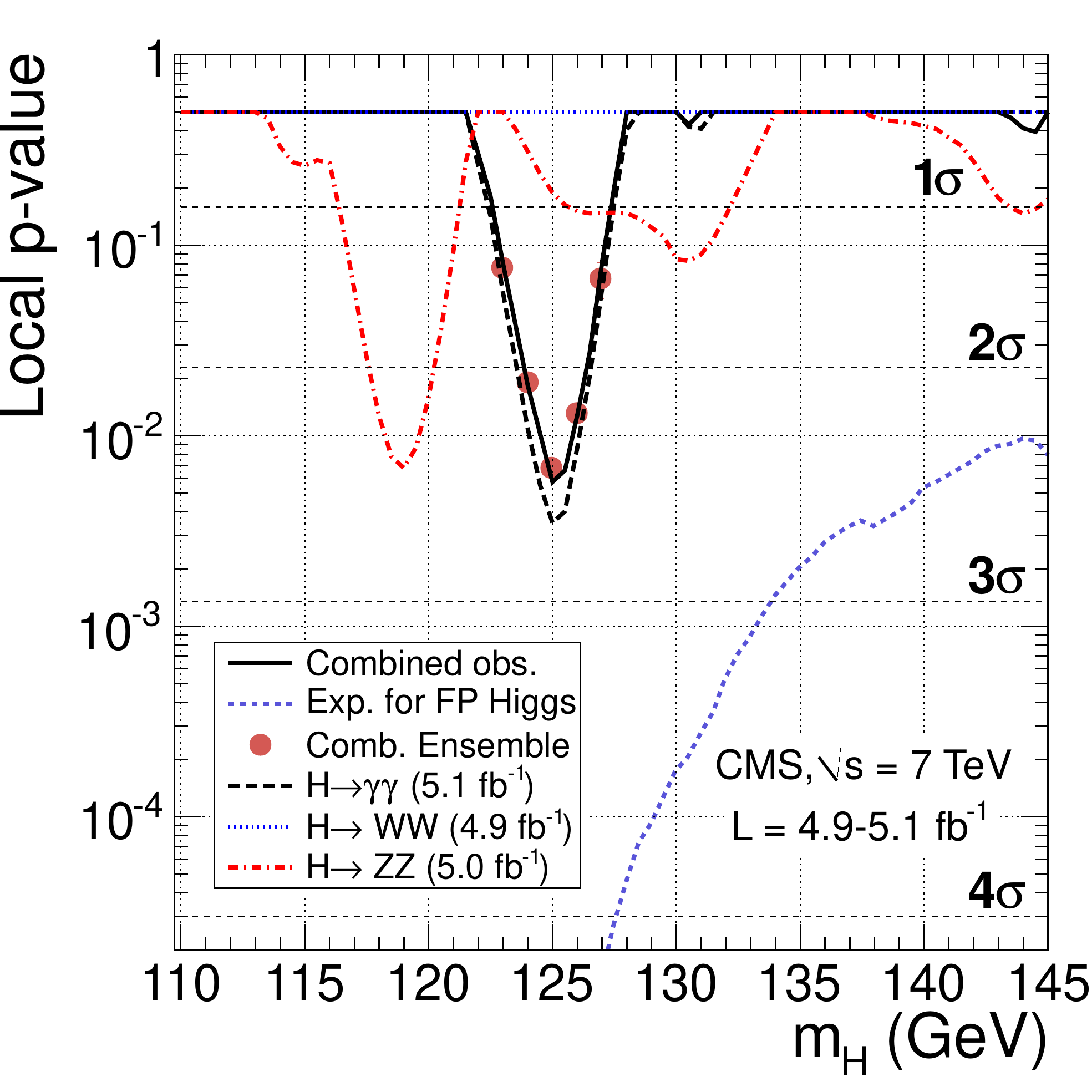} \hfill
\includegraphics[width=0.48\textwidth]{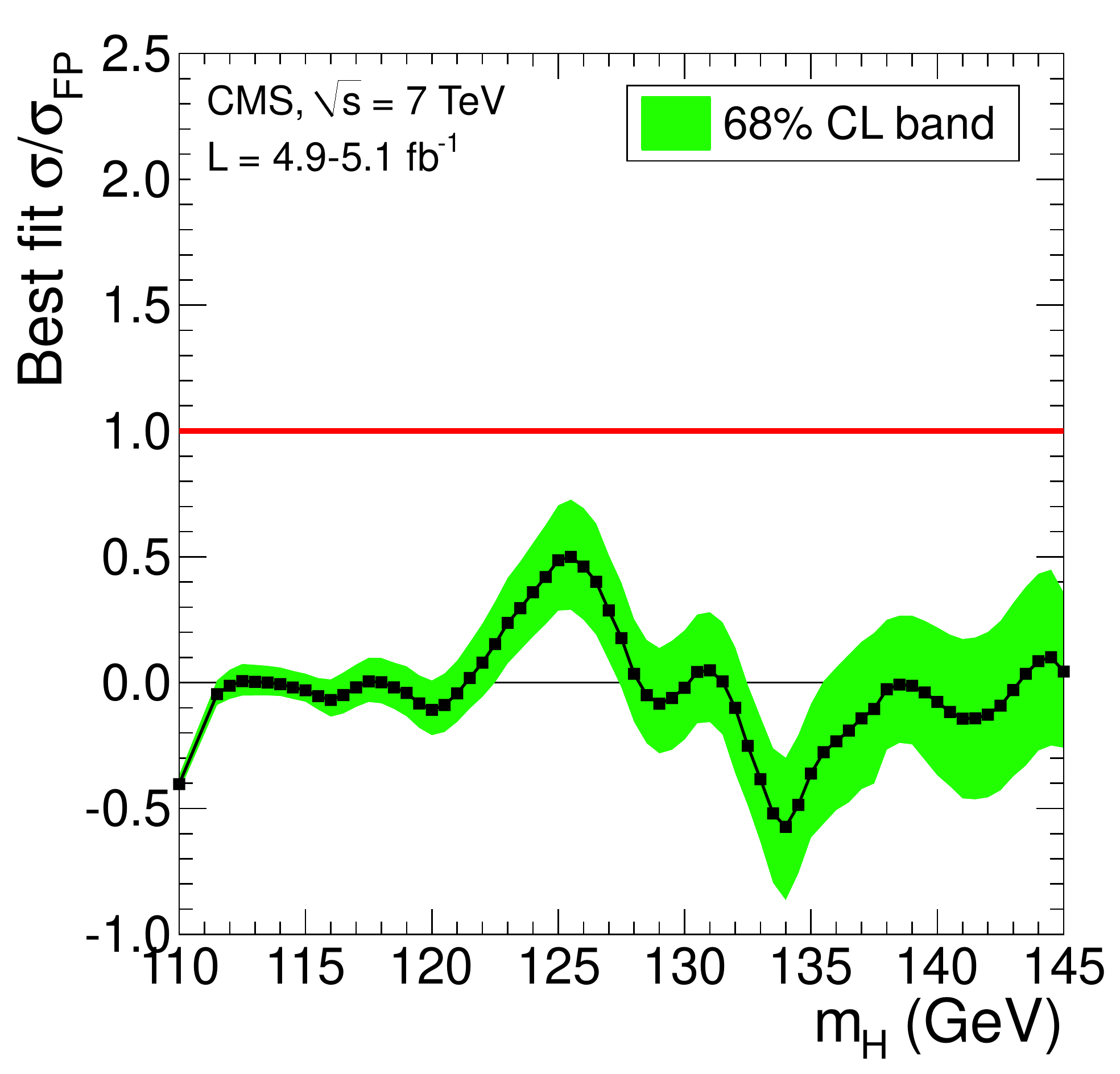}
\caption{
(Left) The observed and expected local $p$-values
as a function of Higgs boson mass for the fermiophobic Higgs boson search.
The local $p$-values for the three decay modes taken separately,
and their combination are obtained with the asymptotic approximation;
the combined local $p$-value is validated at a few $\mH$ (dots) using the
full frequentist approach by generating ensembles of background-only pseudo-datasets.
The dashed line indicates the expected combined local $p$-value, should a fermiophobic Higgs boson with a mass $\mH$ exist.
(Right) The best-fit value of the signal strength, $\sigma /\sigma_{FP}$,
for the combined sub-channels, as a function of
the mass of a fermiophobic Higgs boson.
}
\label{fig:FP_pvalue}
\end{figure*}

The local $p$-value as a function of the Higgs boson mass is obtained using the asymptotic approximation
for individual decay modes and for their combination, and is shown in Fig.~\ref{fig:FP_pvalue} (left).
For a few points, the $p$-value calculation is checked with the frequentist approach~\cite{LHC-HCG} and is shown to agree within the statistical error.
The local $p$-value corresponding to the largest upwards fluctuation
of the observed limit, at \FPMaxZmass\GeV, is computed to have a significance of 2.5~$\sigma$.
When taking into consideration the look-elsewhere effect~\cite{LEE} in the search range 110--300\GeV,
the global significance of the deviation is 0.9~$\sigma$.
This deviation from the expected limit is too weak to be consistent with the fermiophobic Higgs boson signal, as can be seen in Fig.~\ref{fig:FP_pvalue} (right), which shows the observed signal strength for an FP Higgs, as obtained from the fit of signal plus background on data.
In this fit the constraint on signal strength being non-negative is not applied, so that a negative value indicates an observation below the expectation from the background-only hypothesis.

\clearpage
\section{Summary}

Combined results are reported from searches for a fermiophobic Higgs boson
in proton-proton collisions at $\sqrt{s}=7$\TeV
in its decay modes into vector bosons:
$\Pgg\Pgg$, $\PW\PW$, and $\cPZ\cPZ$,
where the Higgs boson production is restricted to vector boson fusion
and associated production with a vector boson.
The Higgs boson mass range explored is 110--300\GeV.
The data analyzed correspond to an integrated luminosity of 4.9--5.1\fbinv.
The fermiophobic Higgs boson is excluded at 95\%~CL
in the mass range \FPObsA --\FPObsB\GeV and
at 99\%~CL in the mass ranges 110--124.5\GeV, 127--147.5\GeV, and 155--180\GeV.

\section*{Acknowledgements}

\hyphenation{Bundes-ministerium Forschungs-gemeinschaft Forschungs-zentren} We congratulate our colleagues in the CERN accelerator departments for the excellent performance of the LHC machine. We thank the technical and administrative staff at CERN and other CMS institutes. This work was supported by the Austrian Federal Ministry of Science and Research; the Belgium Fonds de la Recherche Scientifique, and Fonds voor Wetenschappelijk Onderzoek; the Brazilian Funding Agencies (CNPq, CAPES, FAPERJ, and FAPESP); the Bulgarian Ministry of Education and Science; CERN; the Chinese Academy of Sciences, Ministry of Science and Technology, and National Natural Science Foundation of China; the Colombian Funding Agency (COLCIENCIAS); the Croatian Ministry of Science, Education and Sport; the Research Promotion Foundation, Cyprus; the Ministry of Education and Research, Recurrent financing contract SF0690030s09 and European Regional Development Fund, Estonia; the Academy of Finland, Finnish Ministry of Education and Culture, and Helsinki Institute of Physics; the Institut National de Physique Nucl\'eaire et de Physique des Particules~/~CNRS, and Commissariat \`a l'\'Energie Atomique et aux \'Energies Alternatives~/~CEA, France; the Bundesministerium f\"ur Bildung und Forschung, Deutsche Forschungsgemeinschaft, and Helmholtz-Gemeinschaft Deutscher Forschungszentren, Germany; the General Secretariat for Research and Technology, Greece; the National Scientific Research Foundation, and National Office for Research and Technology, Hungary; the Department of Atomic Energy and the Department of Science and Technology, India; the Institute for Studies in Theoretical Physics and Mathematics, Iran; the Science Foundation, Ireland; the Istituto Nazionale di Fisica Nucleare, Italy; the Korean Ministry of Education, Science and Technology and the World Class University program of NRF, Korea; the Lithuanian Academy of Sciences; the Mexican Funding Agencies (CINVESTAV, CONACYT, SEP, and UASLP-FAI); the Ministry of Science and Innovation, New Zealand; the Pakistan Atomic Energy Commission; the Ministry of Science and Higher Education and the National Science Centre, Poland; the Funda\c{c}\~ao para a Ci\^encia e a Tecnologia, Portugal; JINR (Armenia, Belarus, Georgia, Ukraine, Uzbekistan); the Ministry of Education and Science of the Russian Federation, the Federal Agency of Atomic Energy of the Russian Federation, Russian Academy of Sciences, and the Russian Foundation for Basic Research; the Ministry of Science and Technological Development of Serbia; the Ministerio de Ciencia e Innovaci\'on, and Programa Consolider-Ingenio 2010, Spain; the Swiss Funding Agencies (ETH Board, ETH Zurich, PSI, SNF, UniZH, Canton Zurich, and SER); the National Science Council, Taipei; the Scientific and Technical Research Council of Turkey, and Turkish Atomic Energy Authority; the Science and Technology Facilities Council, UK; the US Department of Energy, and the US National Science Foundation.
Individuals have received support from the Marie-Curie programme and the European Research Council (European Union); the Leventis Foundation; the A. P. Sloan Foundation; the Alexander von Humboldt Foundation; the Belgian Federal Science Policy Office; the Fonds pour la Formation \`a la Recherche dans l'Industrie et dans l'Agriculture (FRIA-Belgium); the Agentschap voor Innovatie door Wetenschap en Technologie (IWT-Belgium); the Council of Science and Industrial Research, India; and the HOMING PLUS programme of Foundation for Polish Science, cofinanced from European Union, Regional Development Fund.

\bibliography{auto_generated}   % will be created by the tdr script.

\cleardoublepage \appendix\section{The CMS Collaboration \label{app:collab}}\begin{sloppypar}\hyphenpenalty=5000\widowpenalty=500\clubpenalty=5000\textbf{Yerevan Physics Institute,  Yerevan,  Armenia}\\*[0pt]
S.~Chatrchyan, V.~Khachatryan, A.M.~Sirunyan, A.~Tumasyan
\vskip\cmsinstskip
\textbf{Institut f\"{u}r Hochenergiephysik der OeAW,  Wien,  Austria}\\*[0pt]
W.~Adam, E.~Aguilo, T.~Bergauer, M.~Dragicevic, J.~Er\"{o}, C.~Fabjan\cmsAuthorMark{1}, M.~Friedl, R.~Fr\"{u}hwirth\cmsAuthorMark{1}, V.M.~Ghete, J.~Hammer, N.~H\"{o}rmann, J.~Hrubec, M.~Jeitler\cmsAuthorMark{1}, W.~Kiesenhofer, V.~Kn\"{u}nz, M.~Krammer\cmsAuthorMark{1}, D.~Liko, I.~Mikulec, M.~Pernicka$^{\textrm{\dag}}$, B.~Rahbaran, C.~Rohringer, H.~Rohringer, R.~Sch\"{o}fbeck, J.~Strauss, A.~Taurok, P.~Wagner, W.~Waltenberger, G.~Walzel, E.~Widl, C.-E.~Wulz\cmsAuthorMark{1}
% ...
\vskip\cmsinstskip
\dag:~Deceased\\
1:~~Also at Vienna University of Technology, Vienna, Austria\\
% ...

\end{sloppypar}
\end{document}